\documentclass{article}
\usepackage{amsthm,amsmath}               
\newtheorem{theorem}{Theorem}

\newtheorem{assumption}{Assumption}
\newtheorem{remark}{Remark}
\usepackage{PRIMEarxiv}

\usepackage[utf8]{inputenc} 
\usepackage[T1]{fontenc}    
\usepackage{hyperref}       
\usepackage{url}            
\usepackage{booktabs}       
\usepackage{amsfonts}       
\usepackage{nicefrac}       
\usepackage{microtype}      
\usepackage{lipsum}
\usepackage{fancyhdr}       
\usepackage{graphicx}       

\graphicspath{{media/}}     

\title{Extended State Observer for Mismatch Disturbances Using Taylor Approximation of the Integral
}

\author{
  Cuong Duc Nguyen \\
  Hanoi University of Science and Technology\\
  Viettel Aerospace Institute\\
  Hanoi \\
  Vietnam\\
  \texttt{cuongnd.192190@sis.hust.edu.vn} \\
}

\begin{document}
\maketitle

\begin{abstract}
The development of disturbance estimators using extended state observers (ESOs) typically assumes that the system is observable. This paper introduces an improved method for systems that are initially unobservable, leveraging Taylor expansion to approximate the integral of disturbance dynamics. A new extended system is formulated based on this approximation, enabling the design of an observer that achieves exponential stability of the error dynamics. The proposed method's efficacy is demonstrated through a practical example, highlighting its potential for robust disturbance estimation in dynamic systems.
\end{abstract}

\keywords{Extended state observer \and Taylor approximation \and Unobservable. \vphantom{\cite{ref1},\cite{ref2},\cite{ref3},\cite{ref4},\cite{ref5},\cite{ref6},\cite{ref7},\cite{ref8},\cite{ref9},\cite{ref10}}}

\section{Introduction}
Disturbance observation is essential in practical control systems where complete model information is unavailable. Disturbance observers find wide applications in controlling systems like robotics \cite{ref1}, flight systems \cite{ref2}, and servo motors \cite{ref3}. Furthermore, the design of state observers underscores the need to observe disturbances affecting the system state. This has spurred extensive research into developing observers that can monitor both the system state and disturbances, commonly known as extended state observers (ESOs). The theory of ESO was initially developed in the early 1970s \cite{ref4}, constructing an extended state vector by incorporating disturbance terms into the state vector and subsequently designing an observer for this augmented system. However, this method is limited to disturbances following linear differential equations. More generally, the limitations of this approach were highlighted in \cite{ref7}, which is only applicable to systems where the disturbances or their higher derivatives are negligible or eliminable. In response, in the study presented in \cite{ref7}, a new ESO based on backward differential approximation was proposed to capture the dynamics of disturbances. This additional dynamic component in the observer is designed to mitigate the influence of disturbance derivatives.

 However, all the aforementioned methods assume that the extended system containing disturbance terms in the state vector is observable. In practice, this condition is often challenging to meet, prompting further investigations into system observability \cite{ref8, ref9}. These studies introduce the concept of strong observability to ensure the convergence of the observer; however, in \cite{ref10}, it was demonstrated that in certain cases, disturbances can still be observed even when the system is only strongly detectable. This relaxation of the strongly observable condition eases the constraints typically encountered in observer design.

Building upon the concepts discussed above, the primary objective of this article is to enhance the standard ESO method for achieving improved estimations of both system state and disturbances. The main contribution of this work is the design of a novel observer tailored for systems with unobservable disturbances, achieved through the use of Taylor expansion to approximate the integral component of disturbances.

This paper is structured as follows: Section 2 presents the general result outlining the proposed methodology. Section 3 illustrates the application of this approach through an example and provides simulation results to demonstrate its effectiveness. Finally, Section 4 concludes the paper with a summary about the work.

\section{PROBLEM STATEMENT}
\label{sec2}
Consider the system following:
\begin{equation}\label{eq1}
\left\{\begin{array}{l}
\dot{x}(t)=A x(t)+B u(t)+d(t)\\
y(t)=C x(t)
\end{array}\right.
\end{equation}
where $x(t) \in \mathbb{R}^{n}, u(t) \in \mathbb{R}^{m }, y(t) \in \mathbb{R}^{p}, d(t) \in \mathbb{R}^{n}, A \in \mathbb{R}^{n\times n},  B \in \mathbb{R}^{n\times m}, C\in \mathbb{R}^{p\times n},p<n$. The matrices $A$, $B$, $C$, $D$ are known constant matrices and control input $u(t)$ is known. The uncertain component $d(t)$ is the combination of disturbances and unmodelled dynamic. The general assumptions about the nominal system (\ref{eq1}) and the disturbance are made as follow:
\begin{assumption}
    The pair of $(A,C)$ is observable.
\end{assumption}
\begin{assumption}
    The first derivative and the second derivative of d(t) are bounded as follows:\\
    \begin{equation}\label{eq2}
\|\dot{d}(t)\| \leq d_{1}
    \end{equation}
    \begin{equation}\label{eq3}
\|\ddot{d}(t)\| \leq d_{2}
    \end{equation}
where $d_1$ and $d_2$ are positive constants.
\end{assumption}
Consider the new system state as $X(t)=[x(t),d(t)]^T$ which transform the system (\ref{eq1}) as following:
\begin{equation}\label{eq4}
\left\{\begin{array}{l}
\dot{X}(t)=\bar{A} X(t)+\bar{B} u(t)+\Gamma_{1}(t)  \\
Y(t)=\bar{C} X(t)
\end{array}\right.
\end{equation}
where $
\bar{A}=\left[\begin{array}{cc}
A & D \\
\mathbf{0}_{n\times n} & \mathbf{0}_{n\times n}
\end{array}\right], \bar{B}=\left[\begin{array}{c}
B \\
\mathbf{0}_{n\times m}
\end{array}\right], \bar{C}=\left[\begin{array}{ll}
C & \mathbf{0}_{p \times n}
\end{array}\right],\Gamma_{1}(t)=[\textbf{0}, \dot{d}(t)]^{T},
D=I
$. $I$ is the identity matrix with dimension of $n\times n$. According to \cite{ref10}, the nominal system (\ref{eq4}) ($\Gamma_{1}(t)=0$) is observable if and only if $rank\left(\begin{bmatrix}
        A-\lambda I&D\\
        C&\mathbf{0}_{p\times n}
    \end{bmatrix}  \right) = 2n
,\forall \lambda \in \mathbb{C}$.
\begin{remark}\label{re1}
The condition of the observability cannot be satisfied when the matrix $C$ has less rows than matrix $A$. Indeed, when 
$p<r$ the rank of the matrix $\begin{bmatrix}
        A-\lambda I&D\\
        C&\mathbf{0}_{p\times n}
    \end{bmatrix} $ is $ \leq n+p$, which 
indicates that the system (\ref{eq4}) is unobservable.
\end{remark}
From Remark \ref{re1}, it is evident that the ESO cannot be applied to system (4). Therefore, in the sequel, this paper proposes a method to augment the available output information of the system to enhance its observability properties.\\
Apply the Taylor expansion around zero for disturbance $d(t)$ gives:
\begin{equation}
   \label{eq5}
   d(t)=d(0)+\dot{d}(t)+\frac{1}{2}\ddot{d}(t)+O\left(d^{(3)}(t)\right)
\end{equation}
With the assumption that $O\left(d^{(3)}(t)\right)=0$, the disturbance $d(t)$ can be approximated by its first and second derivatives. Then using the Taylor expansion (\ref{eq5}), the time derivative of $d(t)$ can be expressed as:
\begin{equation}
    \label{eq6}
\dot{d}(t)=\frac{d(t)-d(t-h)}{h}-R(t, h)
\end{equation}
where $h>0$ is a constant delay that can be tuning and $R(t, h)$ is the remainder of the Taylor approximation \cite{ref7}. The approximation (\ref{eq6}) is also known as backward difference approximation \cite{ref7}. Note that, the remainder component is satisfied the following statement:
\begin{equation*}\label{eq7}
|R(t, h)| \leq \frac{d_{2}}{2} h
\end{equation*}
And the integral of $d(t)$ over an interval of time can also be approximated as follow:
\begin{equation}\label{eq8}
\int_{t-h}^{t} d(t) d\tau= d(t)-d(t-h)+\frac{1}{2h}\left[d(t)+d(t-2h)\right]-\frac{1}{h}d(t-h)+C(t,h)
\end{equation}
where $C(t,h)=d(0)h+R(t-h,h)-R(t,h)$ and $\|C(t,h)\| \leq h\|d(0)\|+d_{2}h $.
\begin{assumption}\label{a3}
    Matrix $A$ is invertible and the smallest eigenvalue $\lambda_{min}\left(A+L\bar{C_2}\right)$ exceeds $k$-time the largest eigenvalue $\lambda_{max}\left(L\bar{C_3}\right)$ with $k>1$ and the observer gain $L$, which will be introduced later. 
\end{assumption}
Thanks to Assumption \ref{a3}, the integral of the output signal over an interval of time can be expressed as:
\begin{equation}\label{eq9}
\int_{t-h}^{t}y(t) d\tau= CA^{-1}\left[x(t)-x(t-h)-B\int_{t-h}^{t}u(t)d\tau - \int_{t-h}^{t}d(t)d\tau \right]
\end{equation}
Inserting equation (\ref{eq8}) to (\ref{eq9}) gives:
\begin{equation}
    \label{eq10}
    \int_{t-h}^{t}W(t)d\tau=CA^{-1}\left[x(t)-x(t-h)-\left(d(t)-\left(1+\frac{1}{h}\right)d(t-h)+\frac{1}{2h}\left[d(t)+d(t-2h)\right]+C(t,h)\right) \right]
\end{equation}
where $W(t)= y(t)+CA^{-1}Bu(t)$.\\
Consider the new output vector $Y=\left[y,\int_{t-h}^{t}W(t)d\tau\right]^T$, the system (\ref{eq4}) is rewritten as follow:
\begin{equation}
    \label{eq10}
\left\{\begin{array}{l}
\dot{X}(t)=\bar{A} X(t)+\bar{B} u(t)+\frac{1}{h} D_{2}[X(t)-X(t-h)]+\Gamma_{2}(t)  \\
Y(t)=\bar{C_2} X(t)+\bar{C_3}X(t-h)+\frac{1}{h}D_4X(t-h)-\frac{1}{2h}D_5[X(t)+X(t-2h)]+\Gamma_3(t)
\end{array}\right.
\end{equation}
with $\Gamma_{2}(t) \in \mathbb{R}^{(n+q) \times 1}$ and $\Gamma_{2}(t)=[\textbf{0},-R(t, h)]^{T},\Gamma_{3}(t)=[\textbf{0},- CA^{-1}C(t, h)]^{T}$ and $
D_{2}=\left[\begin{array}{cc}
\textbf{0} & \textbf{0} \\
\textbf{0} & I_{n}
\end{array}\right]
$, $
\bar{C_2}=\left[\begin{array}{cc}
C & \textbf{0} \\
CA^{-1} & -CA^{-1}
\end{array}\right]
$, $
\bar{C_3}=\left[\begin{array}{cc}
\textbf{0} & \textbf{0} \\
-CA^{-1} & CA^{-1}
\end{array}\right]
$, $D_4=D_5=\left[\begin{array}{cc}
\textbf{0} & \textbf{0} \\
\textbf{0} & CA^{-1}
\end{array}\right]$.
\begin{assumption}
    \label{a4}
    The pair of $(\bar{A},\bar{C_2})$ is observable.
\end{assumption}
Then, we proposed a new observer with an artificial delay as:
\begin{align}\label{eq11}
\dot{\hat{X}}(t)= &\bar{A} \hat{X}(t)+\bar{B} u(t)+\frac{1}{h} D_{2}[\hat{X}(t)-\hat{X}(t-h)]\nonumber\\
&+L[\bar{C_2} \hat{X}(t)-Y(t)]+ L\left[\bar{C}_{3}\hat{X}(t-h)+\frac{1}{h} D_{4}\hat{X}(t-h)-\frac{1}{2h}D_{5} [\hat{X}(t)+\hat{X}(t-2h)]\right]
\end{align}
where $L$ can be chosen such that $\bar{A}+L\bar{C_2}$ is a Hurwitz matrix.\\
Then, the dynamic of the observer error $e(t)=\hat{X}(t)-X(t)$ can be expressed as:
\begin{align}
    \label{eq12}
\dot{e}(t)=&(\bar{A}+L \bar{C}_2) e(t)+\frac{1}{h} D_{2}[e(t)-e(t-h)]\nonumber\\
&+L\bar{C}_{3}e(t-h)+L\frac{1}{h} D_{4}e(t-h)-L\frac{1}{2h}D_{5}[e(t)+e(t-2h)]-\Gamma_2(t)-L\Gamma_3(t)
\end{align}
To analyze the stability condition for (\ref{eq12}), we will proof the theorem below.

\begin{theorem}\cite{ref7} Consider the system (\ref{eq10}) with the observer (\ref{eq11}). There exist $h^{*}, \alpha_{2}, \beta_{2}, \gamma_{2},\gamma_{3}>0$ such that for all $h>h^{*}$, the observation error verifies:

\begin{equation}\label{eq13}
\|e(t)\| \leq \alpha_{2}\|e(0)\| e^{-\beta_{2} t}+\gamma_{2} d_{2} h +\gamma_{3} \|d(0)\| h 
\end{equation}

\end{theorem}
\begin{proof}
    Consider the Lyapunov candidate as follow:
    \begin{equation}
        \label{eq14}
        V=e^T(t)Pe(t)
    \end{equation}
where $P>0, Q>0$ is the solution of Lypunov equation: $
(\bar{A}+L \bar{C}_2) P+P(\bar{A}+L \bar{C}_2)^{T}=-Q $.\\
First we consider ($\Gamma_{2}+L\Gamma_{3}=0$). Taking the time-derivative of (\ref{eq14}), yields:
\begin{align}
    \label{eq15}
\dot{V}(t)=&-e^{T}(t) Q e(t)+\frac{2}{h} e^{T}(t) P D_{2}[e(t)-e(t-h)] +2e^T(t)PL\bar{C}_3e(t-h)\nonumber\\
&+\frac{2}{h} e^{T}(t) P LD_{4}e(t-h)-\frac{1}{h} e^{T}(t) PL D_{5}[e(t)+e(t-2h)]
\end{align}
Then, we have:
\begin{align}
    \label{eq16}
\dot{V}(t) \leq&-c_{3}\|e(t)\|^{2}+c_4\|e(t)\|\|e(t-h)\|+\frac{c_{5}}{h}\|e(t)\|\|e(t)\|+\frac{c_{6}}{h}\|e(t)\|\|e(t-h)\|+\frac{c_{7}}{h}\|e(t)\|\|e(t-2h)\| 
\end{align}
where $c_3 = \lambda_{min}(Q)>0, c_4= \lambda_{max}(2PL\bar{C_3})>0, c_5>0$.\\
From \cite{ref5}, the following condition is assumed:
\begin{equation}\label{eq17}
V(t+\theta) \leq \kappa V(t), \quad \forall \theta \in[-h, 0],\kappa>1
\end{equation}
Thus, one can get:
\begin{equation}\label{eq18}
\|e(t-h)\| \leq c_{8}\|e(t)\| 
\end{equation}
with $c_{8}=\sqrt{\kappa \frac{\lambda_{\max }(P)}{\lambda_{\min }(P)}}$. Utilizing the above inequality, one obtains:
\begin{equation}\label{eq19}
\dot{V}(t) \leq-\left(c_{3}-c_4c_8-\frac{c_{5}}{h}-\frac{c_{6} c_{8}}{h}-\frac{c_{7} c_{8}}{h}\right)\|e(t)\|^{2} 
\end{equation}
Let $c_{9}=c_{3}-c_4c_8-\frac{c_{5}}{h}-\frac{c_{6} c_{8}}{h}-\frac{c_{7} c_{8}}{h}$, one has:
\begin{equation}\label{eq20}
\dot{V}(t, e(t)) \leq-c_{9}\|e(t)\|^{2} 
\end{equation}
Given a sufficiently large value of $h$ and under Assumption \ref{a3}, we can ensure that $c_{9}>0$. Furthermore, integrating the uncertainty components ($\Gamma_{2}+L\Gamma_{3}\neq0)$ into the error dynamics, the comparison method in \cite{ref6} guarantees that inequality (\ref{eq13}) is satisfied.  This completes the proof.
\end{proof}
\begin{remark}
    As per Reference \cite{ref7}, the artificial delay $h$ and the observer gain $L$ can be chosen appropriately to reduce the size of the convergence ball defined by inequality (\ref{eq13}).
\end{remark}

\section{EXAMPLE}
In this section, we will consider the linear system with no input as follow:
\begin{equation}
    \label{eq21}
    \left\{\begin{array}{l}
    \dot{x}(t)=A x(t)+d(t)\\
    y(t)=C x(t)
    \end{array}\right.
\end{equation}
where $A=\begin{bmatrix}
    -1&-2\\
    1&-4
\end{bmatrix}$, $C=\begin{bmatrix}
    1&1
\end{bmatrix}$, $d(t)={\begin{bmatrix}
    D(t)&D(t)
\end{bmatrix}}^T$, $D(t)= step(t-500)(\sin(0.2\pi t)+0.2)(\sin(2\pi t)+0.5)$. It is noteworthy that system (\ref{eq21}) is a system without input signals; however, we can treat this system similarly to cases involving input signals. The initial states and the parameters for the simulation can be chosen as  follows: $x_0={\begin{bmatrix}
    -1&-1
\end{bmatrix}}^T$, $\hat{X}_0={\begin{bmatrix}
    0.2&0.3&-0.1&0.2
\end{bmatrix}}^T$, $h=0.05$,  $L=\begin{bmatrix}
        1.3451  &  1.4595\\
    1.8493  &  2.5647\\
   -0.1343 &  -0.3293\\
   -0.1093 &  -1.0129
\end{bmatrix}$.\\
\begin{figure}
  \centering{\includegraphics[width = 0.55\textwidth]{"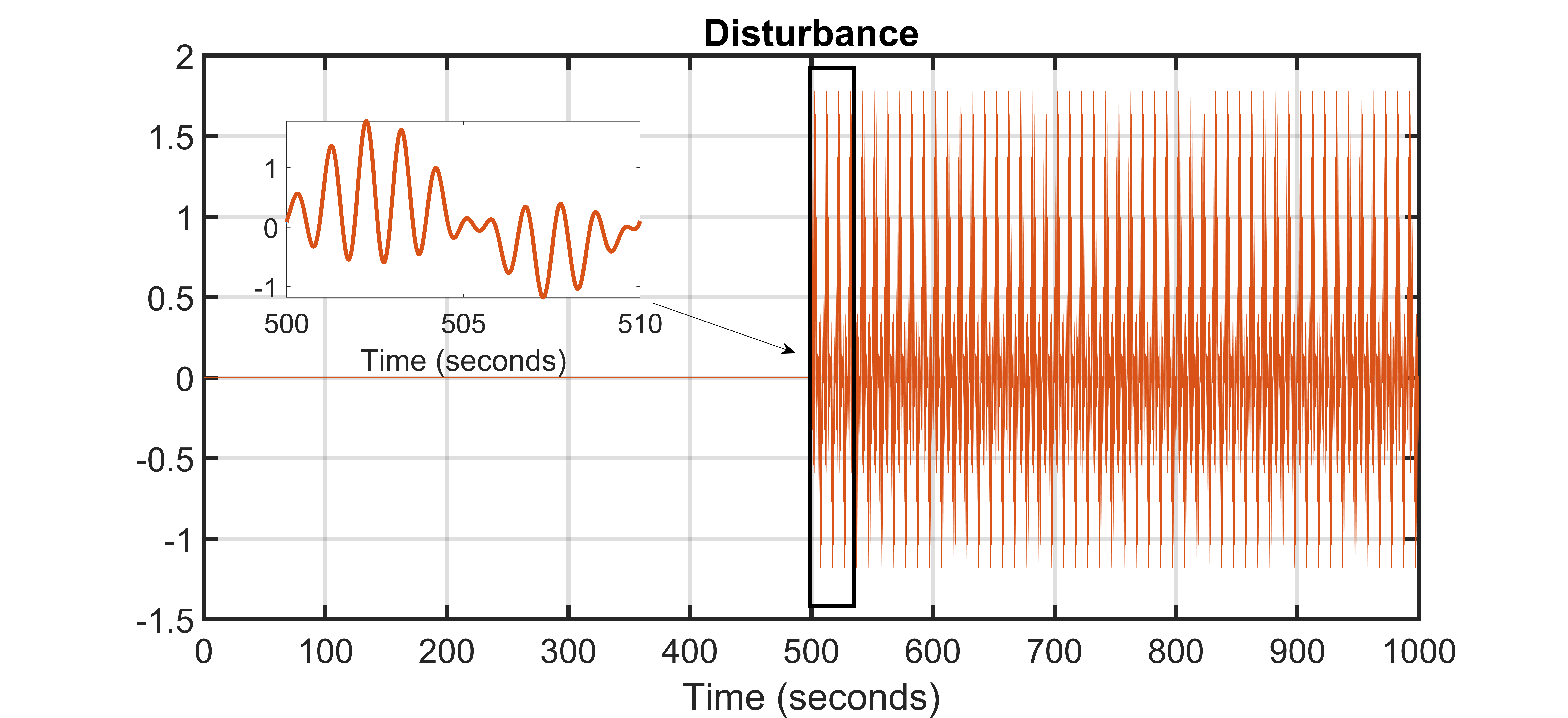"}}
  \caption{Disturbance}
  \label{fig1}
\end{figure}
\begin{figure}
  \centering{\includegraphics[width = 0.55\textwidth]{"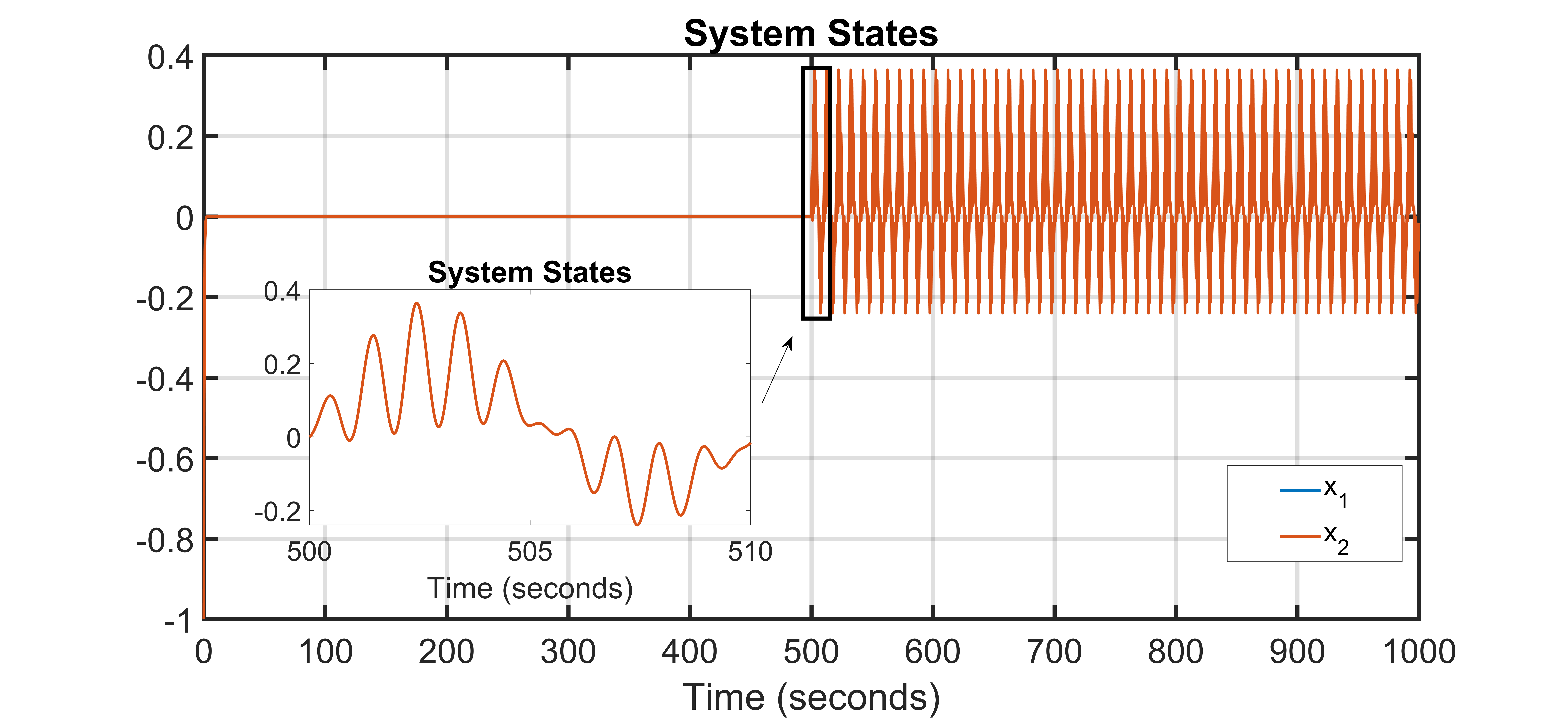"}}
  \caption{System States}
  \label{fig2}
\end{figure}
\begin{figure}
  \centering{\includegraphics[width = 0.55\textwidth]{"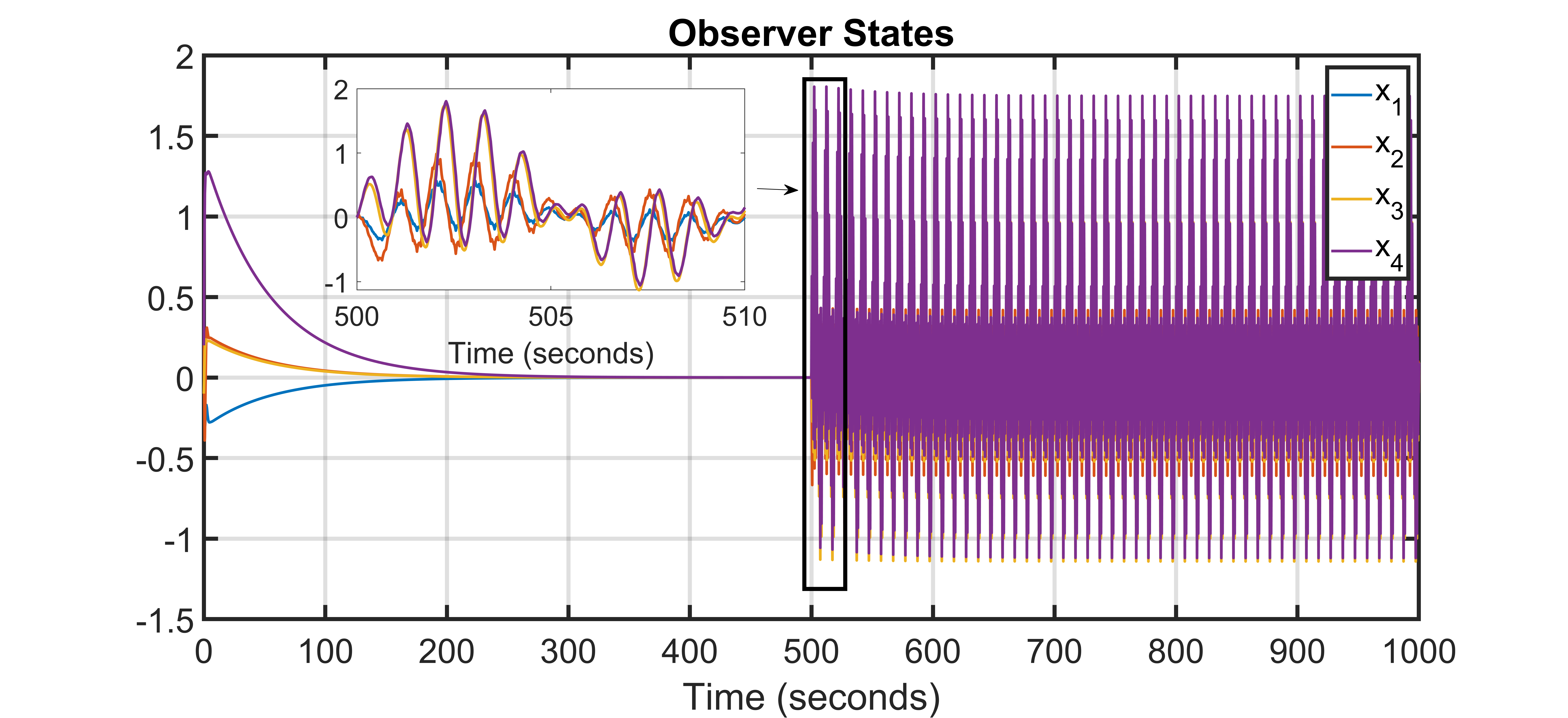"}}
  \caption{Observer States}
  \label{fig3}
\end{figure}
\begin{figure}
  \centering{\includegraphics[width = 0.55\textwidth]{"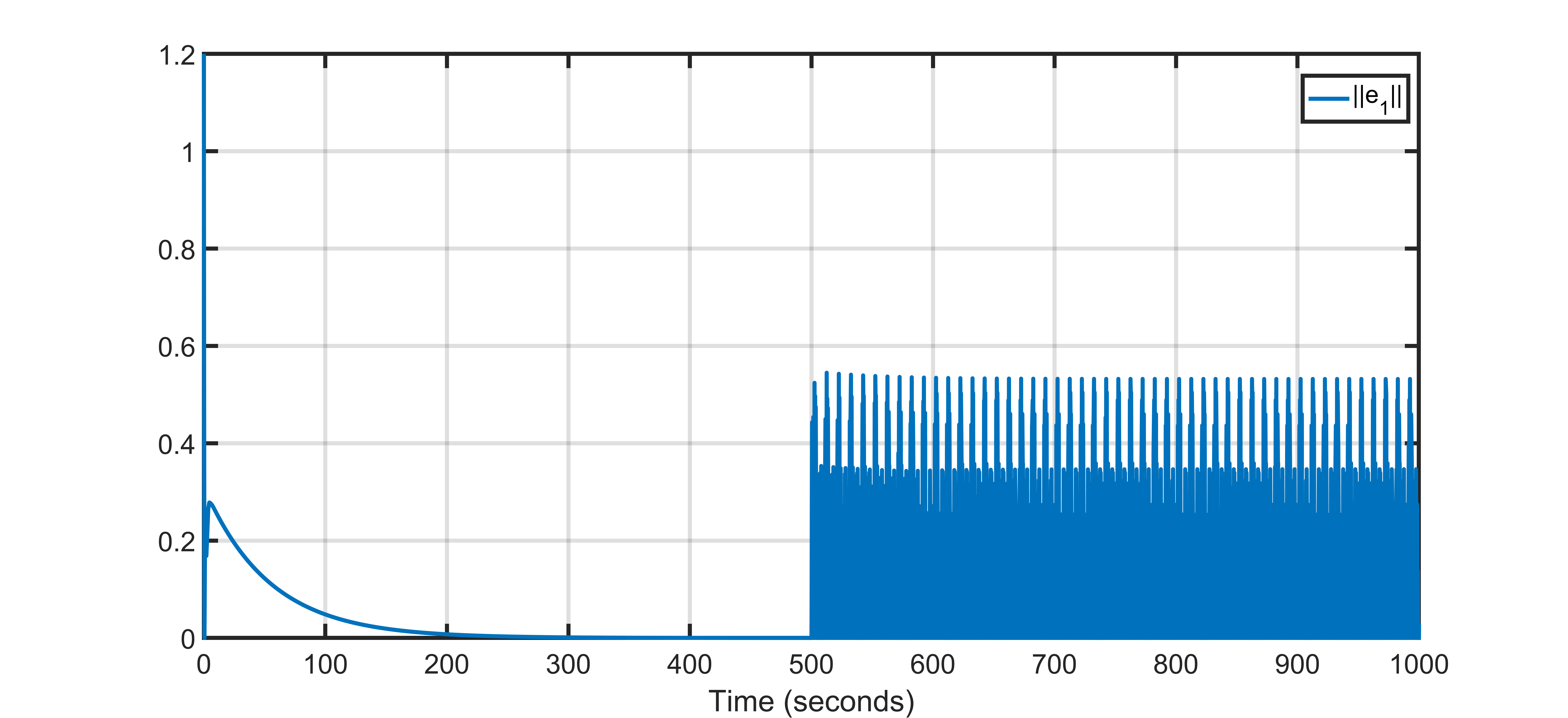"}}
  \caption{Observation error of $x_1$}
  \label{fig4}
\end{figure}
\begin{figure}
  \centering{\includegraphics[width = 0.55\textwidth]{"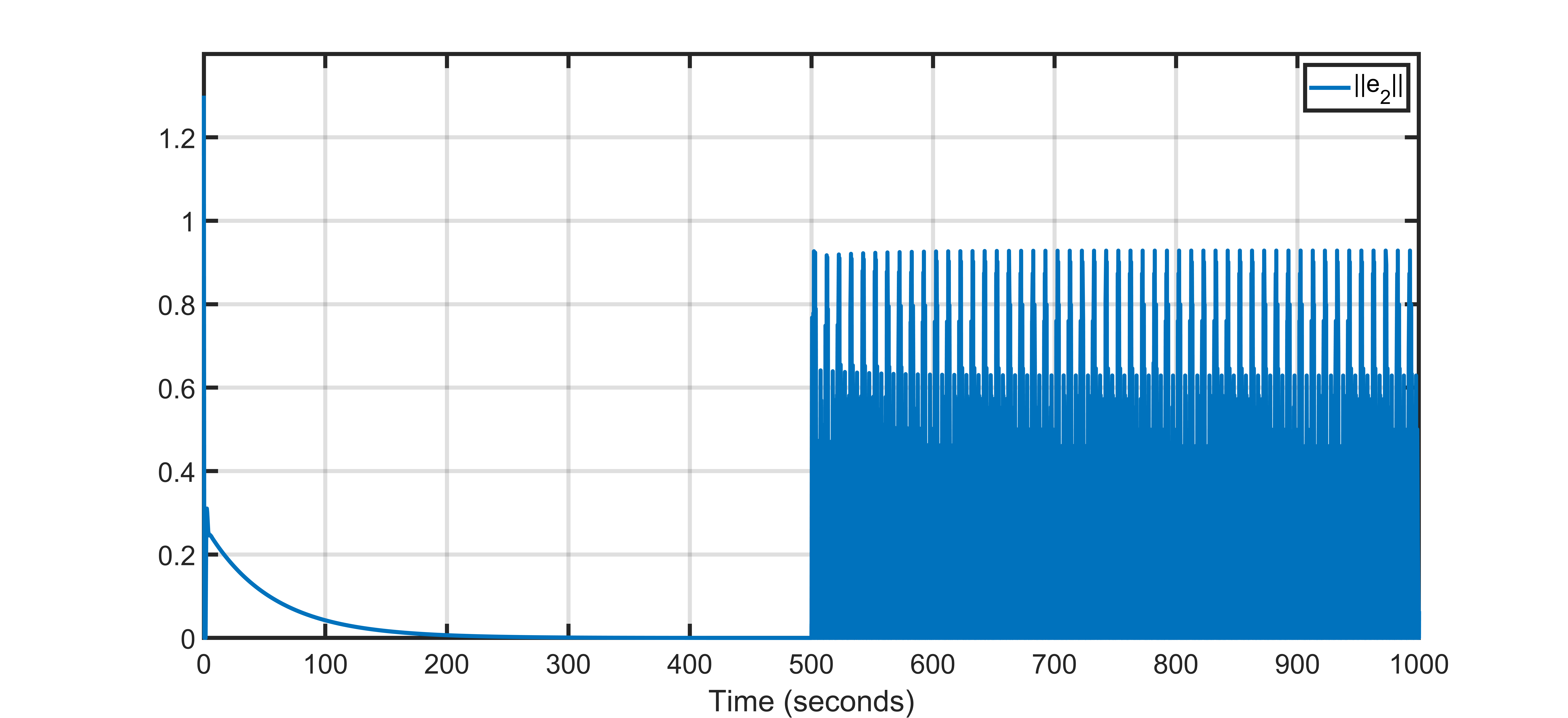"}}
  \caption{Observation error of $x_2$}
  \label{fig5}
\end{figure}
\begin{figure}
  \centering{\includegraphics[width = 0.55\textwidth]{"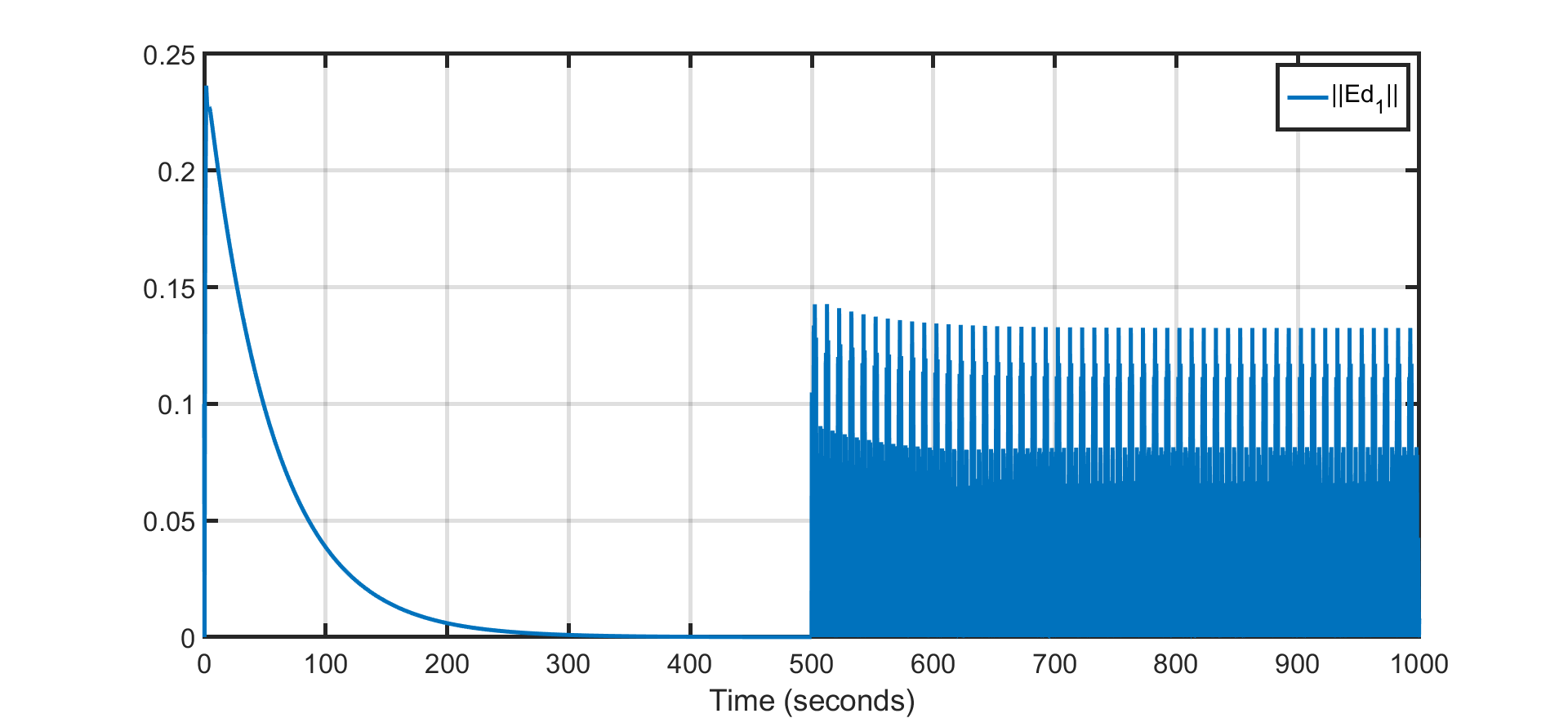"}}
  \caption{Observation error of $d_1$}
  \label{fig6}
\end{figure}
\begin{figure}
  \centering{\includegraphics[width = 0.55\textwidth]{"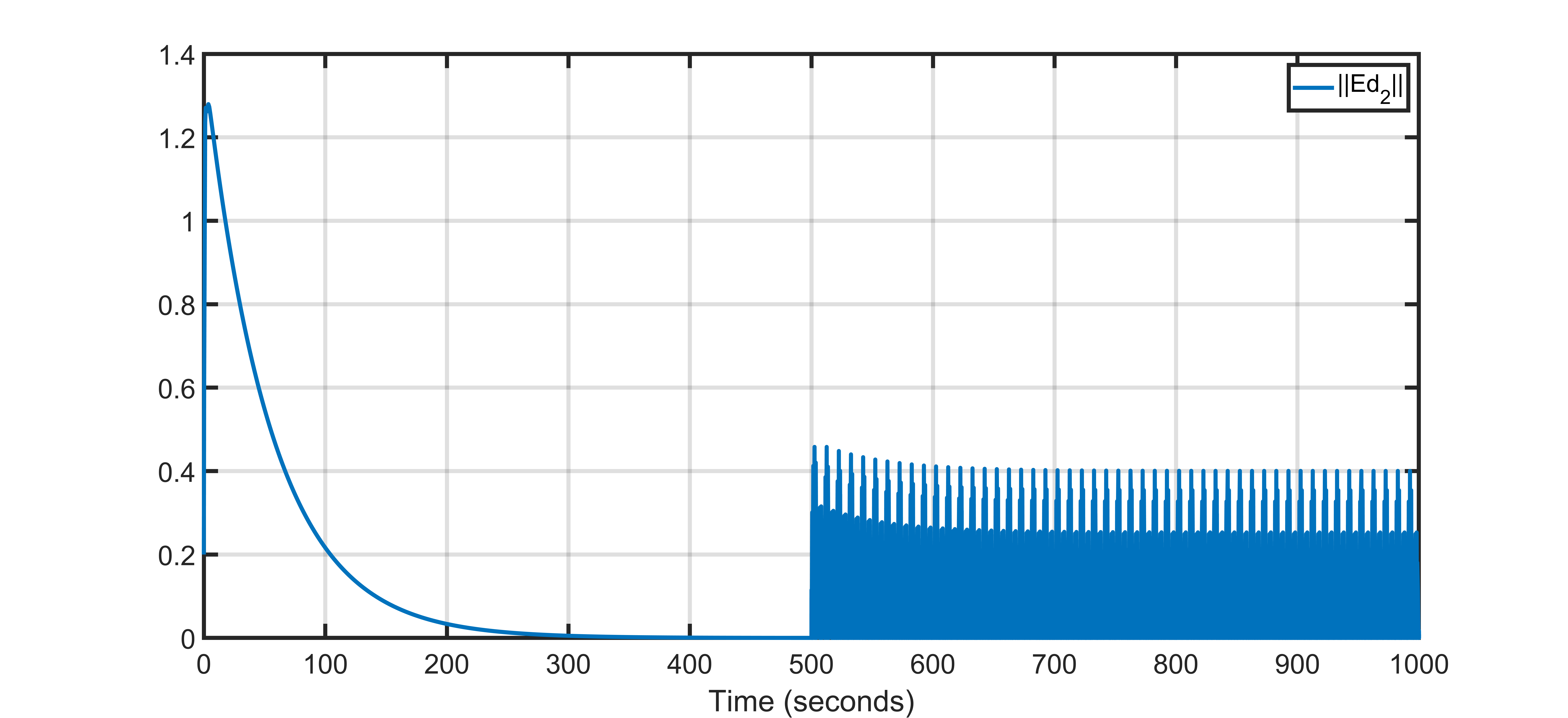"}}
  \caption{Observation error of $d_2$}
  \label{fig7}
\end{figure}
As can be observed in Figure \ref{fig2}, the disturbance depicted in Figure \ref{fig1} impacts the system after 500 seconds, once the system has stabilized. Meanwhile, Figure \ref{fig3} illustrates the system state observation and the effect of noise on the system before and after the time point of 500 seconds. 
In Figures 5 and 6, it is evident that the state observer accurately captured the system state before the noise affected the system at 200 seconds. However, following the impact of the noise, the observed state values were approximately close to the true system values with errors of 0.58 and 0.98 for $x_1$ and $x_2$, respectively.
Additionally, in Figures 7 and 8, it is also evident that the noise signals were accurately observed after 300 seconds, and after the time point of 500 seconds, the observation errors were 0.15 and 0.4 for $d_1$ and $d_2$ respectively.

\section{CONCLUSION}
A new extended system has been proposed for the design of ESO, aiming to improve the precision of observation compared to the standard extended system. This approach involves approximating the disturbance dynamics using a backward difference method and Taylor approximation, which introduces an additional delayed term in the observer. The convergence proof of the estimation error is achieved through a Lyapunov-Razumikhin analysis. It is demonstrated that the accuracy of the estimation is closely tied to the magnitude of the artificial delay. Simulation results highlight the effectiveness of this new design.

\bibliographystyle{unsrt}  
\bibliography{references}

\end{document}